\documentclass[pre,showpacs,floatfix,twocolumn,amsmath,amssymb,superscriptaddress]{revtex4-2}
\usepackage{CJK,graphicx}
\usepackage{color}
\usepackage[colorlinks,bookmarks=false,citecolor=blue,linkcolor=red,urlcolor=blue]{hyperref}
\usepackage{multirow}

\usepackage{ulem}
\graphicspath{{figures/}}

\newcommand{\be}{\begin{equation}}
\newcommand{\ee}{\end{equation}}

\begin{document}
\begin{CJK*}{UTF8}{gbsn}
\title{Surface critical properties of the three-dimensional clock model}
\author{Xuan Zou}
\altaffiliation{The two authors contributed equally to this work.} 
\affiliation{Department of Physics, Beijing Normal University, Beijing 100875, China}
\affiliation{Institute for Advanced Study, Tsinghua University, Beijing 100084, China} 
\author{Shuo Liu}
\altaffiliation{The two authors contributed equally to this work.} 
\affiliation{Department of Physics, Beijing Normal University, Beijing 100875, China}
\affiliation{Institute for Advanced Study, Tsinghua University, Beijing 100084, China} 

\author{Wenan Guo}
\email{waguo@bnu.edu.cn}
\affiliation{Department of Physics, Beijing Normal University, Beijing 100875, China}
\affiliation{Beijing Computational Science Research Center, Beijing 100193, China}

\begin{abstract}

Using Monte Carlo simulations and finite-size scaling analysis, 
we show that the $q$-state clock model with $q=6$ on the simple cubic lattice with open surfaces 
has a rich phase diagram; in particular, it has an extraordinary-log phase, besides
the ordinary and extraordinary transitions at the bulk critical point.
We prove numerically that the presence of the intermediate extraordinary-log phase 
is due to the emergence of an O(2) symmetry in the surface state 
before the surface enters the 
$Z_{q}$ symmetry-breaking region as the surface coupling is increased at the bulk critical point, while O(2) symmetry emerges for the bulk. 
The critical behaviors of the extraordinary-log transition, as well as the ordinary and the special transition separating the ordinary
and the extraordinary-log transition are obtained.

\end{abstract}

\pacs{to be added } 

\maketitle

\section{Introduction}

Surface effects are significant at the bulk critical point due to divergence of the correlation length \cite{Cardy:383311}. The 
surfaces display critical phenomena that differ from the bulk, and one bulk universality class may display several surface universality classes. 
The surface critical behavior (SCB) has attracted numerous investigations
\cite{binder,Diehl}. 
Typically, the rich and novel SCBs include the ordinary, extraordinary and special transitions \cite{PhysRevE.72.016128, PhysRevE.73.056116, PhysRevB.84.134405}. If the coupling in the surface layer $J_s$ is comparable to the bulk coupling $J$, the surface remains disordered before the bulk orders at $T_c$ due to fewer coordinate numbers. 
The surface singularities at the bulk $T_c$ are purely induced by the bulk criticality. This is called `ordinary transition'. 
When the surface coupling $J_s$ is sufficiently enhanced, i.e., $J_s \gg J$, a phase transition can occur on the open surface at $T>T_c$,
which is coined as the surface transition, which belongs to the $D-1$ dimensional universality class of the $D$ dimensional model considered.  
At the bulk transition point, the ordered surface exhibits extra singularities, such a transition is called `extraordinary transition'.  
At a fine-tuned surface coupling strength $J_s^*$, the surface transition temperature and the bulk $T_c$ coincide, forming a multi critical point 
called `special transition'. 

Recently, the research on SCBs has attracted renewed attention in the quantum spin models, where novel results are reported.\cite{zhanglong,Ding2018,Weber1,weber2,zhu2021}
A novel extraordinary-log universality class has been proposed for a surface critical state in 
three-dimensional (3D) O$(N)$ models by Metlitski \cite{metlitski2020boundary}, with $2\le N < N_c$.
This extraordinary-log SCB has been verified in classical O(3) $\phi^4$ model \cite{PhysRevLett.126.135701} and 3D $XY$ model \cite{PhysRevLett.127.120603},
i.e., there is such an extraordinary-log phase when $J_{s}$ is sufficiently enhanced.
A significant important and exciting direction is to identify whether other related models display this novel logarithmic surface universality class. 

In this work, we study the surface critical behavior of the 3D $q$-state clock model with $q>4$.
At the bulk critical point, the O(2) symmetry emerges. The transition, therefore, belongs to the 3D O(2) universality class.
The ordinary SCB associated is expected naturally in the same class as the 3D $XY$ model. 

However, different from the 3D $XY$ model, the surface is in the $Z_q$ symmetry broken phase
at large $J_s$. This is due to the fact that,
at the large $J_s$ limit, the surface is described by the two-dimensional (2D) $q$-state clock model, which is $Z_q$ ordered at low temperature.
Therefore, there is an extraordinary phase transition in the clock model, where the $Z_q$ ordered surface obtained extra singularities due to bulk critical fluctuations.

The 2D $q$-state clock model is rather special which has two `melting' temperatures, $T_1$ and $T_2$, when $q>4$ \cite{PhysRevB.16.1217, PhysRevB.33.437}.  
The system is in a symmetry broken phase if temperature $T<T_1$. When $T_1\le T \le T_2$, the system melts to a critical 
Berezinskii-Kosterlitz-Thouless (BKT) phase \cite{Kosterlitz1973, berezinskii1971destruction, berezinskii1972destruction}. 
Further increasing temperature to $T>T_2$, the system completely melts to a high-temperature disordered phase. 
In the BKT phase, where the O(2) symmetry emerges, there is no long-range order, but the correlation decays in a power law with exponent $\eta$ varying between $\eta(T_1)=4/q^2$ and $\eta(T_2)=1/4$ \cite{PhysRevB.16.1217}. 
More detailed discussions can be found in \cite{nienhuis1984critical}. 
Therefore, in the case of the surface coupling $J_s \gg J$, we expect the surface of the 3D clock model undergoes two 
surface transitions
described by the 2D $q$-state clock model at temperatures higher than $T_c$.

\begin{figure}[!h]
\includegraphics[width=1 \columnwidth]{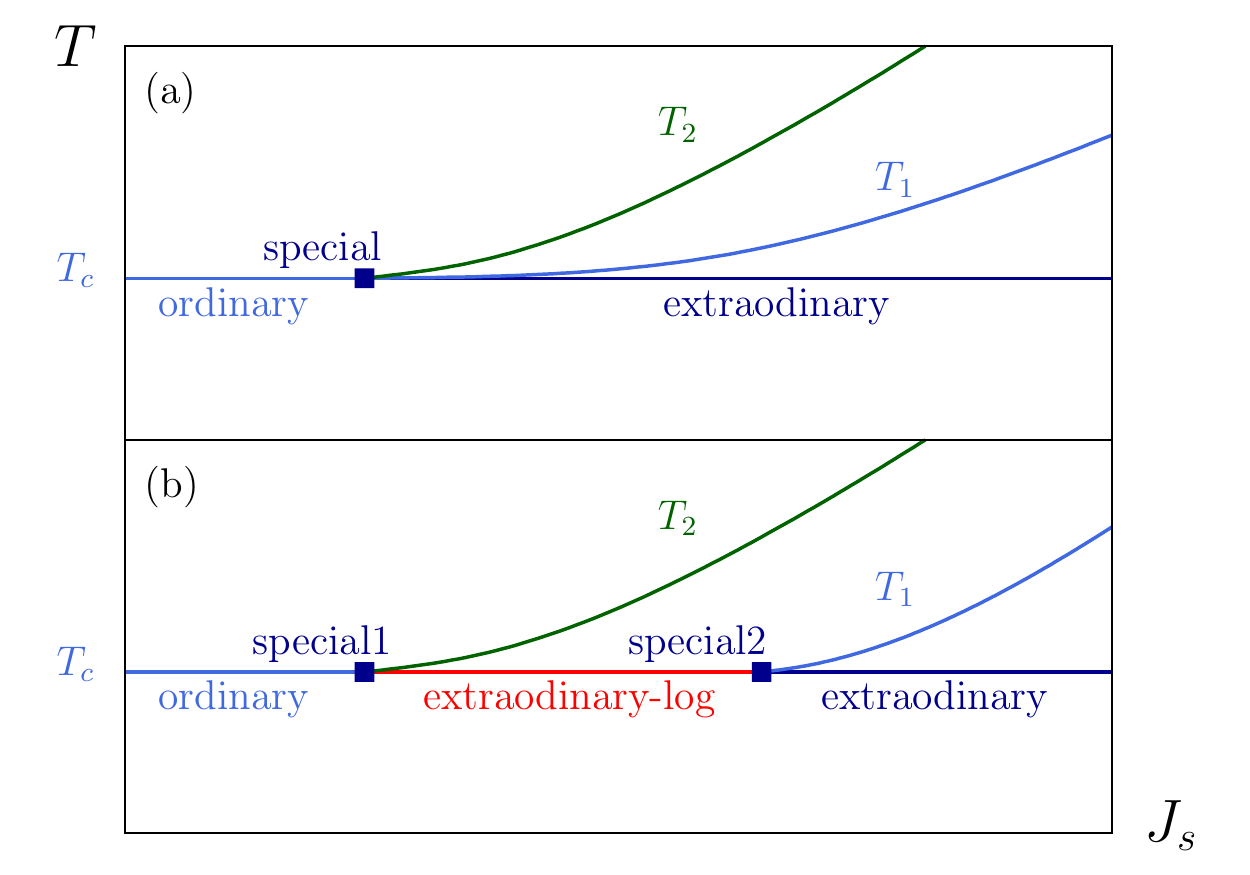}
	\caption{The two possible scenarios for the schematic phase diagram of the 3D 6-state clock model with open surfaces, in which $T_c=2.202$ is the bulk critical temperature, $J_s^*=1.622$ is the first special transition point. Based on the numerical results in the main text, there is an extraordinary-log phase between the ordinary phase and the extraordinary phase, as shown in (b).
\label{phasediagram}}
\end{figure}

Now consider decreasing $J_s$, it is natural to expect that $T_2$ and $T_1$ decrease with $J_s$. The question is: will the temperatures $T_2$ and $T_1$ finally merge at a fine tuned $J_s^*$, which is a multi critical special transition point, as depicted in Fig. \ref{phasediagram}(a), or will the two temperatures decrease to the bulk critical
temperature $T_c$ at two different $J_s$, i.e.,  $J^*_{s}$ and $J_{s}^\dagger$, which we dub as special point 1 and special point 2, respectively, 
as shown in Fig. \ref{phasediagram}(b)?
If the latter case is true, then a question following naturally is: what is the nature of the surface phase between the two 
special points?

In this paper, we address these issues by studying the SCBs and surface transitions in the $q$-state clock model (specifically $q=6$)
using Monte Carlo simulations. We show that it is the second scenario realized. 
We demonstrate numerically that an O(2) symmetry emerges at intermediate $J_s$. This suggests that the emergent O(2) symmetry in the surface at large $J_s$ persists when $T_2$ and $T_1$ decrease to the bulk critical point $T_c$ at $J^*_{s}$ and $J^\dagger_{s}$, respectively, %at bulk critical point $T_c$ 
as illustrated in Fig. \ref{phasediagram}(b).  
We further show that, as the result of this emergent symmetry, 
the intermediate region, between the extraordinary phase and the ordinary phase,
is in the extraordinary-log phase \cite{metlitski2020boundary}.
We determine the first special phase transition point between the ordinary and the extraordinary-log phases. 
The critical properties of the ordinary transition and the first special transition are obtained. 
The results show that the ordinary transition and the special transition are in the same SCB class as the 3D O(2) model, respectively. 
We also verified that, at a large enough surface coupling $J_s$, the surface is $Z_q$ ordered, showing SCBs of extraordinary type. 

The paper is organized as follows:
In Sec. \ref{Sec:model} we introduce the model and observables.
We study the ordinary transition in Sec. \ref{ord} and show the existence of an extraordinary transition in Sec. \ref{extraord}. The two surface transition lines are also revealed at large $J_s$. In Sec. \ref{sec:O2}, we determine the special transition away from the ordinary region and show the existence of an emergent O(2) in the intermediate region between the ordinary region and the extraordinary region. The universal properties of the special transition and the extraordinary-log transition are obtained.
We conclude in Sec. \ref{conclusions}.

\section{Model and Observables}
\label{Sec:model}
We consider a 3D ferromagnetic classical $q$-state clock model on the simple cubic lattice. The Hamiltonian reads
\begin{equation}
H=-\sum_{\langle i,j\rangle} J_{i,j}\cos(\theta_i-\theta_j), %-h\sum_i\cos(q\theta_i),
\label{model}
\end{equation}
in which $\langle i, j \rangle$ denotes nearest-neighbor spins with $J_{i,j}>0$ and $\theta_i=n 2\pi/q, n=1, 2, \cdots,q$. Based on 
previous studies \cite{PhysRevB.61.3430, PhysRevLett.99.207203, PhysRevB.91.174417, PhysRevLett.115.200601, PhysRevB.91.104411, PhysRevB.94.104402, PhysRevE.68.046107, PhysRevB.84.125136},
the phase transition for $q \textgreater 4$  belongs to the 3D O(2) universality class, meaning that the $Z_q$ anisotropy  is irrelevant at the transition point, 
while, for $T<T_c$, the $Z_q$ anisotropy becomes relevant, bringing the O(2) symmetry of the order parameter down to a $q$-fold cyclic permutation 
symmetry $Z_q$. Therefore, the $Z_q$ anisotropy is a prototype of the so-called dangerously irrelevant perturbations (DIP). This model and related
models  have been intensively studied for exploring the physics of DIPs \cite{ShaoPRL}. In present work, we study the 3D 6-state clock model. The bulk critical point is known as $T_c=2.202$ \cite{ShaoPRL}.

To explore the surface critical behaviors of the model, we apply the Wolff cluster Monte Carlo simulations \cite{Wolff}. 
We simulate systems with periodic boundaries along the $x$ and $y$ directions and open boundaries in the $z$ direction. The strength of couplings between spins on the surface layers is denoted as $J_s$, and the strength of other couplings is denoted as $J$. We set $J=1$ throughout the work.

For a given spin configuration, the squared surface magnetization is defined as:
\be
m_s^2= \Big[\frac{1}{L^2}\sum_{x,y} \cos(\theta_{x,y,z})\Big]^2 + \Big[\frac{1}{L^2}\sum_{x,y} \sin(\theta_{x,y,z})\Big]^2,
\ee
where $z=1$, meaning the spin on the top surface, and $L$ is the length of the cubic. It is closely related to the surface susceptibility $\chi_s$,
\be
\chi_s= L^2 m_s^2/T.
\ee
 
The spin correlation functions $C_\parallel(L/2)$ and $C_\perp(L/2)$ are also
adopted to characterize the SCB,
\be
\begin{split}
C_\parallel(L/2)=\frac{1}{2L^2}\sum_{x,y} \langle & \cos (\theta_{x,y,1} - \theta_{x+L/2,y+L/2,1}) \\
+&\cos (\theta_{x,y,L} - \theta_{x+L/2,y+L/2,L}) \rangle,
\end{split}
\ee

\be
C_\perp(L/2)=\frac{1}{L^2}\sum_{x,y} \langle  \cos (\theta_{x,y,1} - \theta_{x,y,L/2}) \rangle.
\ee
We also compute the Binder ratio of surface magnetization 
\be
Q_{s} = \frac{\langle m_s^2\rangle^2}{\langle m_s^4\rangle}.
\ee

%%%%%%%%%%%%%%%%%%%%%%%%%%%%%%%%%%%%%%%%%%%%%%%%%%%%%%%%%%%%%%%%%%%%%%%%%%%%
\section{Ordinary transition}
\label{ord}
We first study the surface state in the region $J_s \approx J$. This is 
the region we expect ordinary transition. We simulate the model at $J_s=1$
in particular.

\begin{figure}[!h]
\includegraphics[width=0.49 \columnwidth]{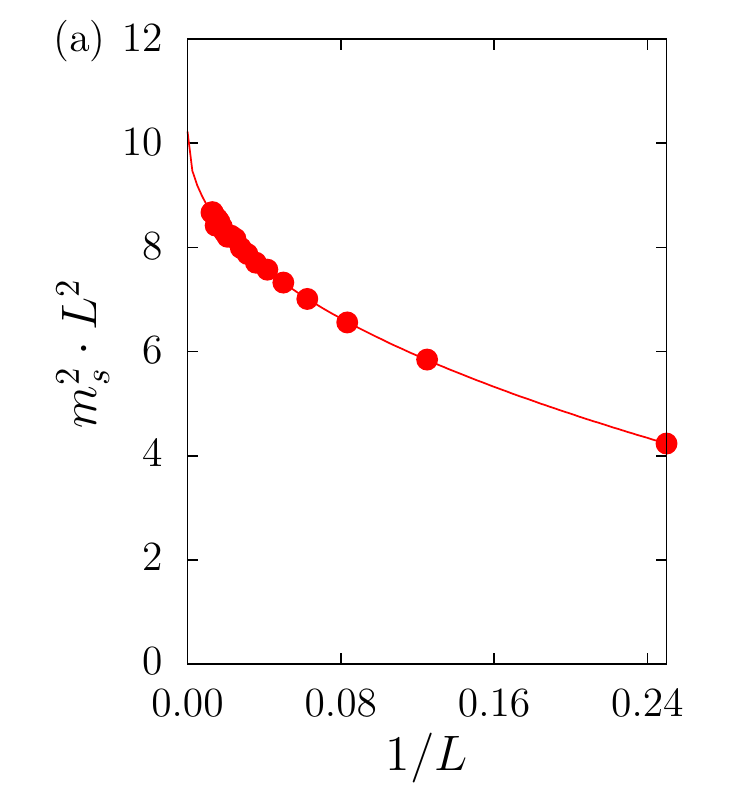}
\hspace{-0.45cm}
\includegraphics[width=0.49 \columnwidth]{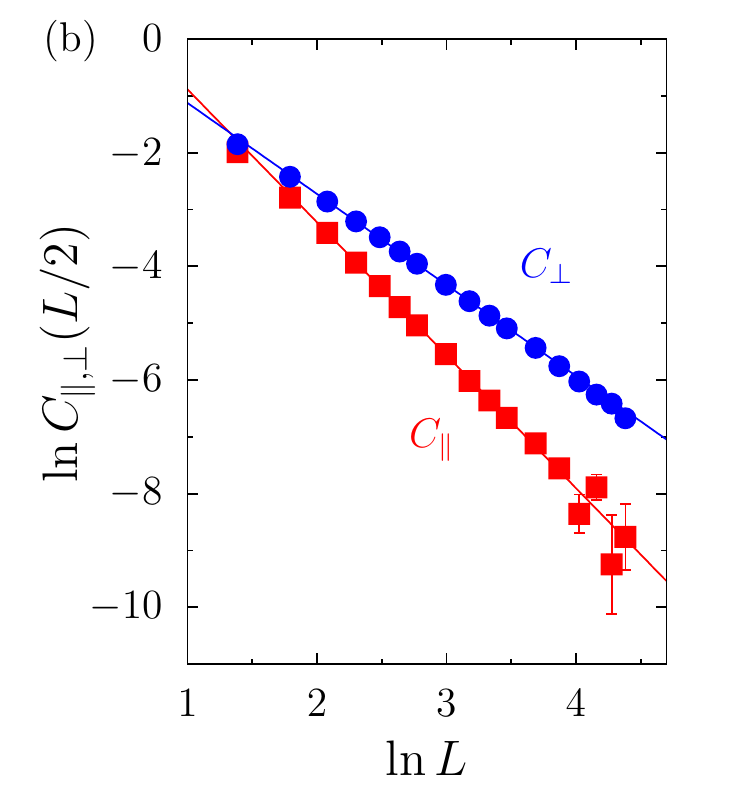}
	\caption{(a), the size-dependence of  $m_s^2L^2$ for $J_s=1.0$; (b), correlation $C_\parallel$ and $C_\perp$ for  $J_s=1.0$ at the ordinary phase transition. The largest size is $L=80$. The lines are the fitting functions. 
\label{ordin}}
\end{figure}

The surface susceptibility $\chi_s$, proportional to $m_s^2L^2$, satisfies the following scaling form for the ordinary transition at bulk $T_c$
\be
	m_{s}^2 L^2= c +a L^{2y_{h1}^{(\rm o)}-2},
	\label{ms2ordin}
\ee
with $c$ an non-universal constant, and $y^{(\rm o)}_{h1}$ the scaling dimension of the surface magnetic field $h_1$. 

Our data for $m_s^2L^2$ are shown in Fig. \ref{ordin}(a) with $J_s=1$.
Fitting Eq. (\ref{ms2ordin}) to the data of $m_s^2L^2$, we obtain $y_{h1}^{(\rm o)}= 0.775(5)$, which agrees with the value for the 3D $XY$ model \cite{PhysRevE.72.016128}.

At ordinary transition, the surface spin correlation at the largest distance $L/2$ obeys the following finite-size scaling forms \cite{PhysRevB.9.2194, PhysRevE.72.016128}
\begin{equation}
	C_\parallel(L/2) = a L^{-(1+\eta^{(\rm o)}_\parallel)},
	\label{cord1}
\end{equation}
and
\begin{equation}
	C_\perp(L/2) = a L^{-(1+\eta^{(\rm o)}_\perp)},
	\label{cord2}
\end{equation}
in which $\eta^{(\rm o)}_{\parallel}$ and $\eta^{(\rm o)}_\perp$ are the surface anomalous exponents. Fig. \ref{ordin}(b) graphs finite-size dependence of $C_\parallel(L/2)$ and $C_\perp(L/2)$ at $J_s=1.0$ in logarithmic scales. The expected power law behaviors are evident. Fitting the data with to the expected forms, Eqs. (\ref{cord1}) and (\ref{cord2}) at $J_s=1.0$,  we find $\eta_\parallel^{(\rm o)}=1.42(6)$ and $\eta_\perp^{(\rm o)}=0.74(1)$. These exponents are listed in Tab. \ref{tab_ord}.

\begin{table}[!h]
    \caption{Surface critical exponents at ordinary  transitions. The exponents of the 3D $XY$ model \cite{PhysRevE.72.016128} are also listed for comparison.}
    \begin{tabular}{ccccc}
        \hline
        \hline
        Type      &   $J_s $  &     $y^{(\rm o)}_{h1}$      & $\eta^{(\rm o)}_\parallel$      & $\eta^{(\rm o)}_{\perp}$   \\
        \hline
        3D clock  &    1.0    &      0.775(5)           & 1.42(6)                     &  0.74(1)               \\
        3D $XY$     &           &      0.781(2)           &                             &                        \\
       \hline
       \hline
    \end{tabular}
    \label{tab_ord}
\end{table}
		
The obtained surface critical exponents $y^{(\rm o)}_{h1}, \eta^{(\rm o)}_\parallel$ and $\eta^{(\rm o)}_\perp$ obey the  scaling relations
\be
\eta_\parallel=d -2 y_{h1},
\label{sc1}
\ee
and
\be
2 \eta_\perp= \eta_\parallel + \eta_b,
\label{sc2}
\ee
in which $\eta_b=0.03853(48)$ \cite{MCeta} is the bulk anomalous dimension for the 3D O(2) universality class.

%%%%%%%%%%%%%%%%%%%%%%%%%%%%%%%%%%%%%%%%%%%%%
\section{Extraordinary transition}
\label{extraord}
\begin{figure}[!h]
\includegraphics[width=0.49 \columnwidth]{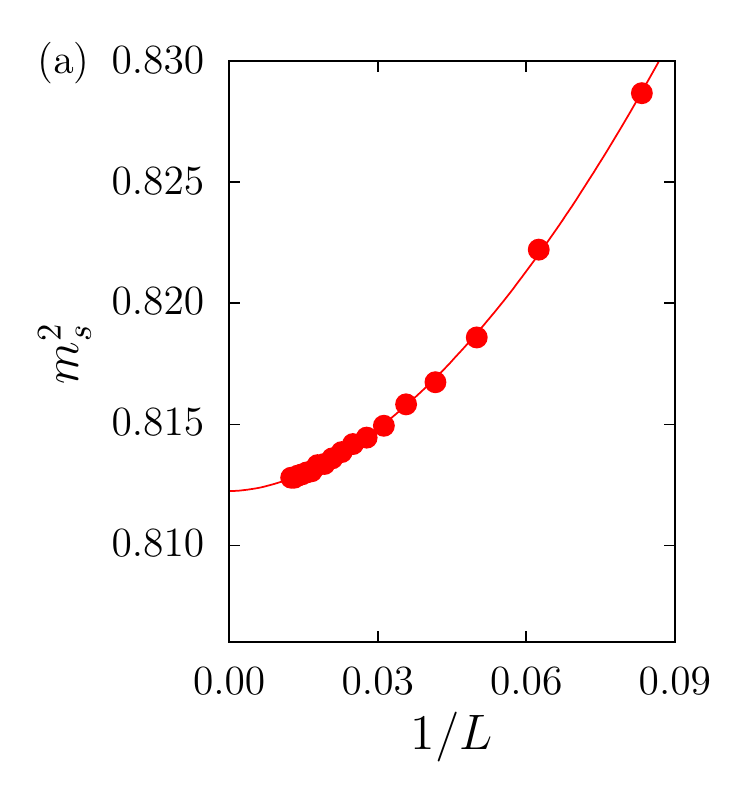}
\hspace{-0.45cm}
\includegraphics[width=0.49 \columnwidth]{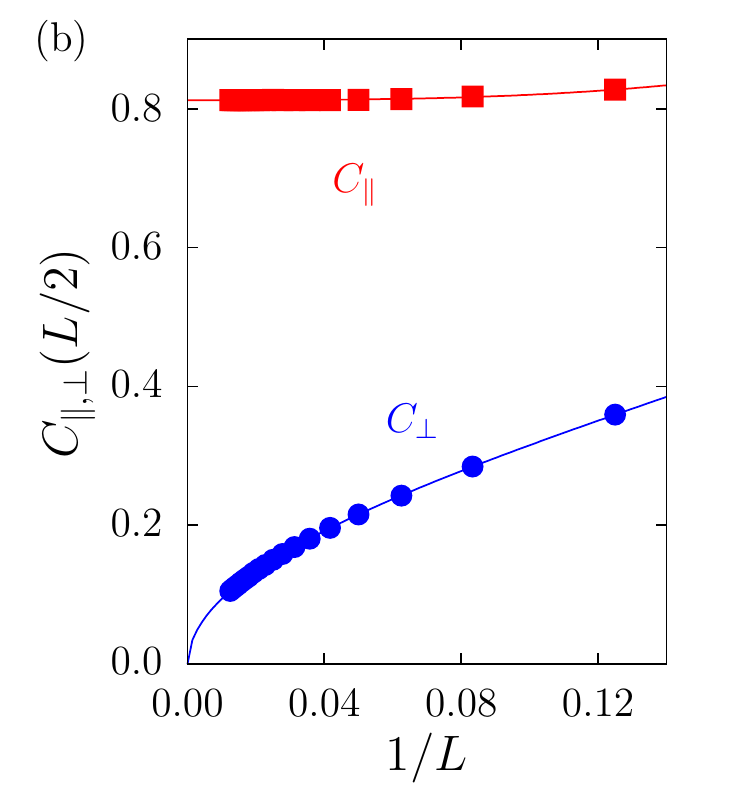}
	\caption{(a), $m_s^2$ and (b), correlation function $C_{\parallel, \perp}$ in the extraordinary phase with $J_s=3.5$. The largest size is $L=80$.
\label{extra}}
\end{figure}

Now let us consider the limit that the surface coupling $J_{s}$ is much larger than the bulk coupling $J$. At this limit, the surface orders before the bulk orders when the temperature is lowered. The surface is described by the 2D $q$-state clock model on the square lattice. At the bulk critical temperature $T_{c}$, we calculate $m_s^2$, $C_{\parallel}$, and $C_{\perp}$ for several $J_s$.  As shown in Fig. \ref{extra}, we find $m_s^2$ and $C_{\parallel}$ converge to finite values at $J_s=3.5$ as system size goes to infinity, which indicates that the surface truly orders with $Z_{q}$ symmetry broken.

It is expected that $m_s^2$ and $C_\parallel$ obtain additional singularities due to bulk critical fluctuations, and scale with $L$
in the following ways
\begin{equation}
	C_\parallel(L/2) = C_0 +a L^{-(1+\eta_\parallel^{(\rm e)})}+\cdots,
	\label{cext}
\end{equation}
\begin{equation}
	m_{s}^2(L) = m^2_0 +a L^{2y_{h1}^{(\rm e)}-4}+\cdots,
	\label{ms2ext}
\end{equation}
with $\eta_\parallel^{(\rm e)}$ and $y_{h1}^{(\rm o)}$ two exponents associated to the extraordinary transition. Here `$\cdots$' means contributions due to analytic
terms. 

We have fitted our data according to Eqs. (\ref{cext}) and (\ref{ms2ext}).
We find $C_0=0.81225(2)$ and $m_0^2=0.81224(4)$.
Unfortunately, we obtain the power ${2y_{h1}^{(\rm e)}-4}$ very close to -2 and $-(1+\eta_\parallel^{(\rm e)})$ close to -1. This makes it difficult to separate the singular parts from the analytic terms.  

Meanwhile, we find $C_{\perp}$ converges to zero, as shown in Fig. \ref{extra}(b). According to theory, it should scale in the following way
\begin{equation}
	C_\perp(L/2) = a L^{-(1+\eta_\perp^{(\rm e)})}+\cdots,
	\label{cordin}
\end{equation}
with $\eta_\perp^{(\rm e)}$ another exponent of extraordinary transition.
Due to the same reason discussed above, we could not obtain this exponent in a reliable value.

When $q>4$, the 2D $q$-state clock model has two `melting' temperatures, $T_1$ and $T_2$, where the transitions are of the BKT type with exponent $\eta$ varying between $\eta(T_1)=4/q^2$ and $\eta(T_2)=1/4$ \cite{PhysRevB.16.1217}.
These correspond to two surface transitions $T_2(J_s)$ and $T_1(J_s)$ at large $J_s$.
To determine the two transition points, we make use of the finite-size scaling of $m_s^2(L)$ 
at the two BKT transitions, which assumes the following form \cite{Kosterlitz1974, PhysRevB.55.3580, PhysRevB.65.184405}
\be
m_s^2(L,T) \propto L^{-\eta} f(\xi/L)
\ee
with $\eta=1/4$ at $T_2$ and $\eta=1/9$ at $T_1$.

\begin{figure}[!h]
\includegraphics[width=0.95 \columnwidth]{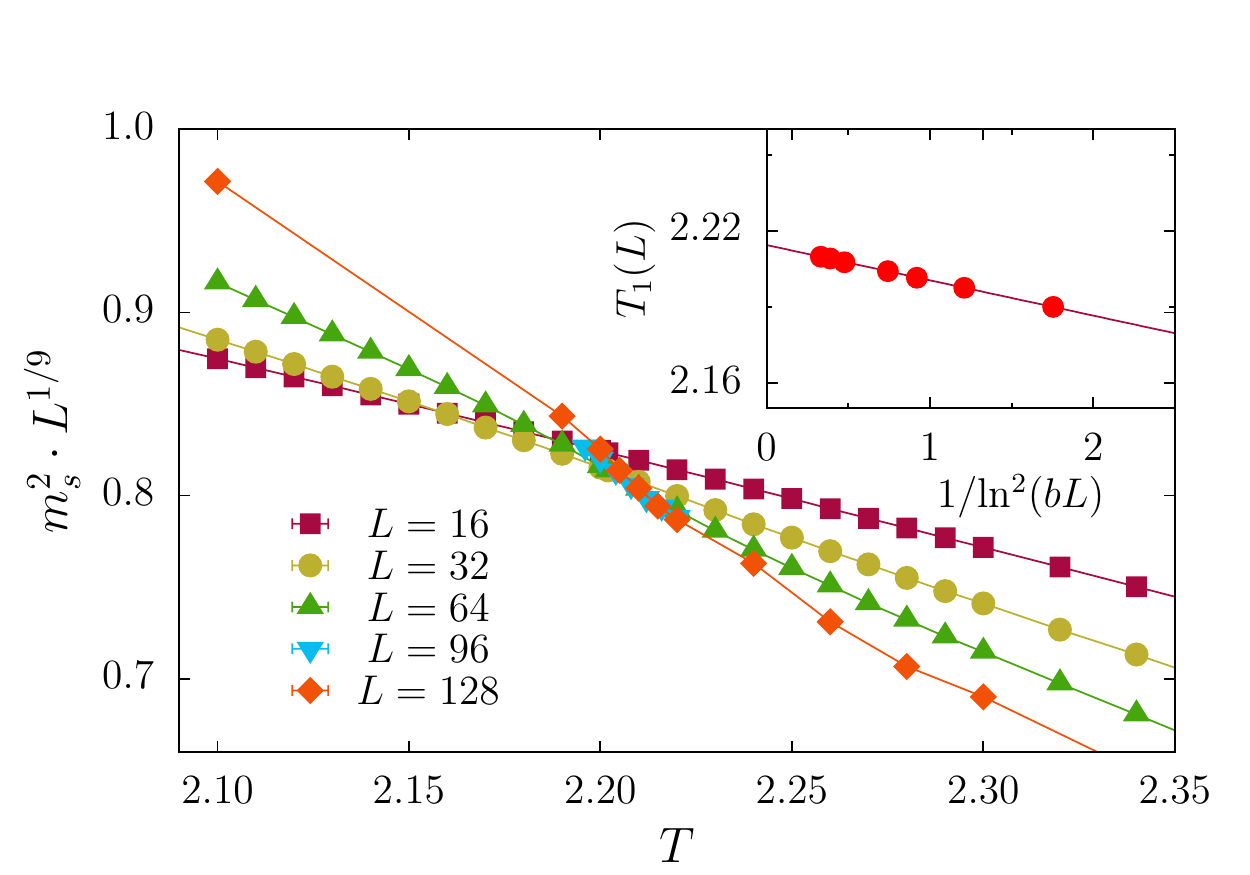}\vspace{-0.5cm}
\includegraphics[width=0.95 \columnwidth]{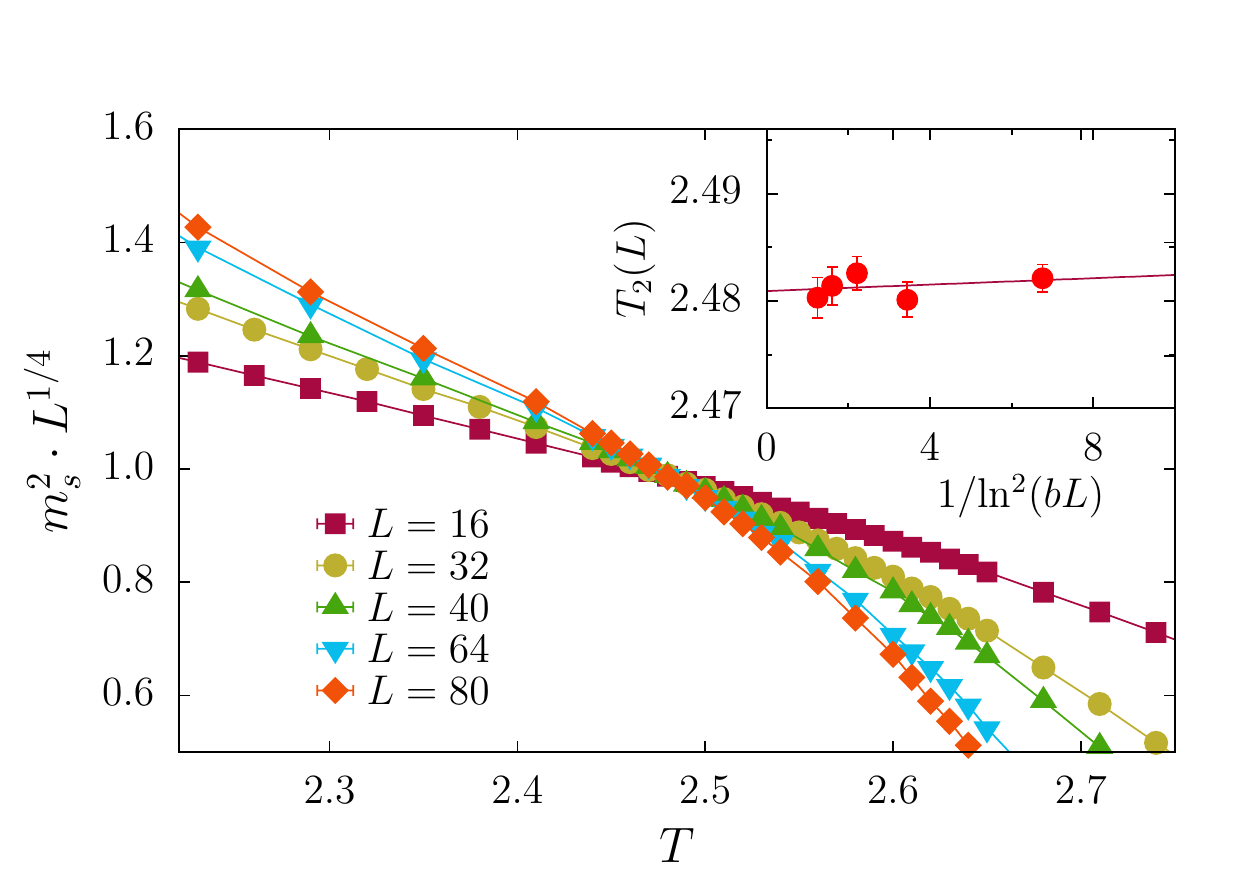}
\caption{The scaled surface magnetization $m_{s}^{2}L^{\eta}$ vs $T$ at $J_s=2.5$. The finite-size estimate of the critical temperature $T(L)$ drafts toward the thermodynamic critical point.}
\label{T2T1b}
\end{figure}

Note that $m_s^2L^{\eta}$ is dimensionless, and it does not change with $L$ at the transition point, which means the curves for various system sizes $L$ should cross at the transition point. 
The finite-size estimate of the critical temperature, $T_2(L)$, corresponds to the crossing point of $m_s^2L^{\eta}$ for $L$ and $2L$, which drifts toward the thermodynamic critical point $T_2$ in the following way for sufficiently large size $L$
\be
T_2(L)-T_2 \propto \frac{1}{\ln^2(bL)}
\ee
with an unknown positive exponent $b$. 
In such a way, we find $T_2=2.843(3)$  for $J_s=3.0$.  
Applying this method to $T_{1}$ and we find $T_1=2.298(2)$. 
Results for other values of $J_s$ are listed in  Tab. \ref{jsT2T1}. 
And this finite temperature BKT transition is shown in Fig. \ref{phasediagram}.
However, for $J_s=2.0$, we find it is very difficult to abstract $T_1$
reliably. 

\begin{table}[h]
\caption{$T_2$ and $T_1$ for various surface coupling $J_s$. }
\begin{tabular}{ccccc}
\hline
\hline
$J_s$&  $T_1$      &$T_2$       \\
\hline
2.5   &  2.2144(5) 	    	&2.481(2)            \\
3.0   &  2.298(2)  	        &2.843(3)                \\
3.5   &  2.60(2)   	        &3.25(1)               \\
\hline
\hline
\end{tabular}
\label{jsT2T1}
\end{table}

A straightforward question is: as we decrease $J_s$,   will $T_1(J_s)$ decrease to $T_c$ of the bulk transition first while $T_2(J_s)$ still higher than $T_c$? 
In between of $T_2$ and $T_1$ the 2D surface acquires an emergent O(2) symmetry. If such emergent O(2) symmetry persists at $T_c$, the surface will
be O(2) symmetric in a region of $J_s$. As a result of such an O(2) symmetry, it is reasonable to expect 
an intermediate extraordinary-log phase between the ordinary phase and the extraordinary phase.

%%%%%%%%%%%%%%%%%%%%%%%%%%%%%%%%%%%%%%%%%%%%%%%%%%%%%%%%%%%%%%%%%%%%%%%%%%%%%%%%%%%%%%%%%%%%%%%%%%%%%%%%%%%%%%%%%%%%%%%%%

%%%%%%%%%%%%%%%%%%%%%%%%%%%%%%%%%%%%%%%%%%%%%%%%%%%%%%%%%%%%%%%%%%%%%%%%%%%%%%%%%%%%%%%%%%%%%%%%%%%%%%%%%%%%%%%%%%%%%%%%%
\section{Emergent O(2) symmetry and Extraordinary-log transition}
\label{sec:O2}
In this section, we will first determine the special transition  leaving the
ordinary phase when $J_s$ is increased. Then, we will show that, in a 
region of $J_s$ larger than the special transition point, the surface has 
an emergent O(2) symmetry, and the phase is controlled by an extraordinary-log fixed point.

\subsection{The first special transition}
\label{subsec:special1}

\begin{figure}[!h]
\includegraphics[width=0.95 \columnwidth]{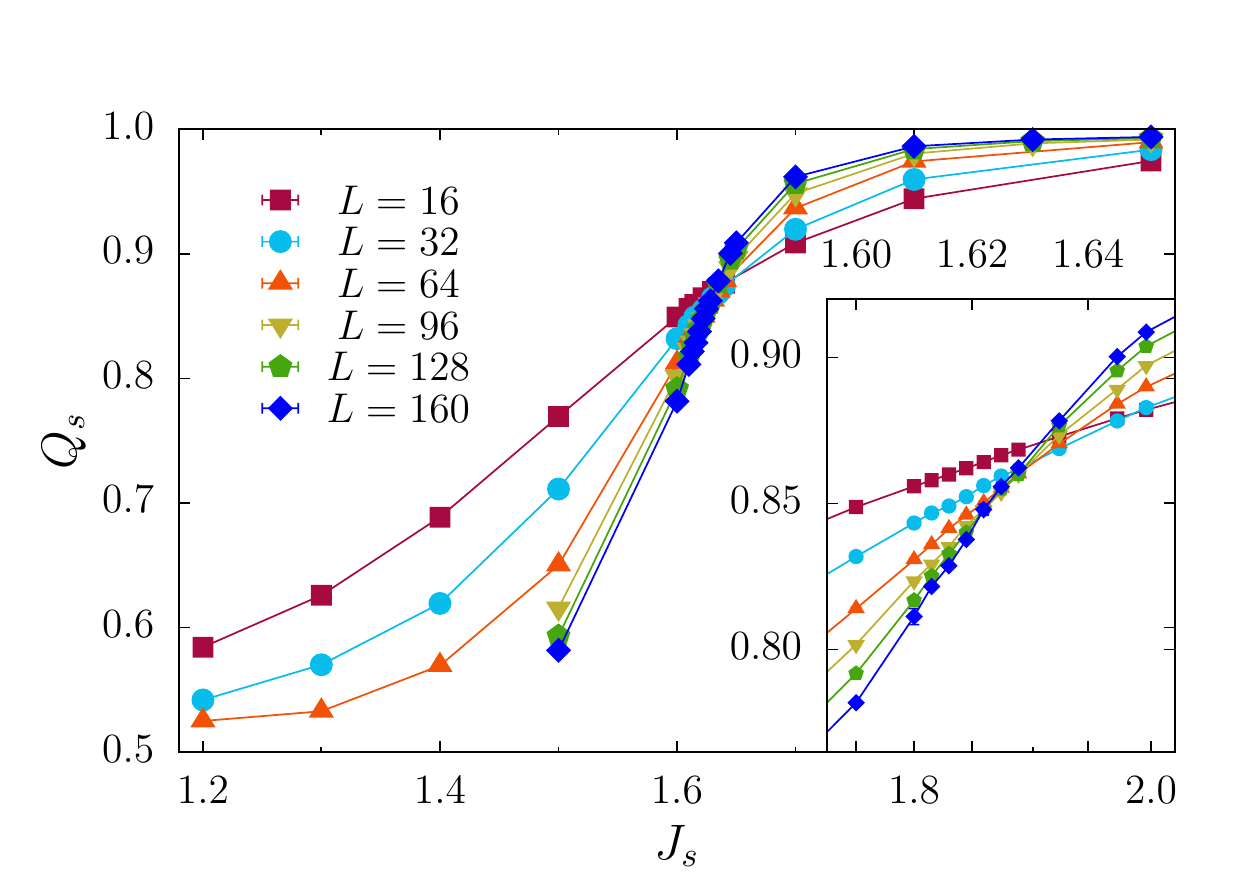}
	\caption{The Binder ratio of the surface magnetization $Q_{s}(T_c, J_s, L)$ for different system sizes $L$ at bulk critical point $T_c$ against surface coupling $J_s$. 	
\label{Jsc}}
\end{figure}

To specify the special transition point,  we calculated the Binder ratio of the surface magnetization $Q_{s}$ at bulk critical point $T_c$.  Part of the data for $Q_s$ 
are graphed against the surface coupling $J_s$  in Fig. \ref{Jsc}. The ratios for different system sizes develop a common crossing point as $L$ increases, which is a standard signal of a phase transition. 

At bulk critical point $T=T_c$ and the vicinity of the fine-tuned special transition point $J_s^*$, the surface scaling field $t_1\propto J_s-J_s^*$ is the only relevant field, therefore,  $Q_s$ has the following finite-size scaling behavior \cite{PhysRevE.72.016128}
\be
Q_s(T_c, J_s, L)= f(T_c, t_1 L^{y_t^{(\rm s)}}),
\ee
in which $y_t^{(\rm s)}$ is the surface thermal exponent associate of $t_1$ and $f$ is a scaling function. If this is true, then 
the crossing point $J_s^*(L)$ of $Q_s$ for $L$ and $2L$ is the finite-size estimate of  $J_s^*$, which converges to the thermodynamic limit value in a power law \cite{doi:10.1126/science.aad5007}
\be
J_s^*(L)-J_s^*\propto  L^{-(y_t^{(\rm s)}+\omega)},
\label{jsfss}
\ee   
where %$y_t^{(\rm s)}$ is the surface thermal exponent, 
$\omega>0$ is the effective exponent of the leading correction to scaling.
Extrapolation using Eq. (\ref{jsfss}) yields  $J_{s}^*=1.622(1)$. We can also fit  the following expansion to the data for $Q_s$ around $J_s^*$
\be
\begin{split}
Q_s(J_s,L)=Q_{s}^{(\rm s)} 
+ a_1(J_s-J_s^*) L^{y_{t1}^{(\rm s)}} +a_2 L^{-\omega}\\
+ a_3(J_s-J_s^*)^2 L^{2 y_{t1}^{(\rm s)}} +a_4 (J_s-J_s^*)  L^{y_{t1}^{(\rm s)}-\omega},
\end{split}
\ee
%where $y_t^{(\rm s)}$ is the surface thermal exponent, and $-\omega<0$ describes the effective leading correction to scaling.
with $a_i (i=1,2,3,4)$ unknown constants, and $Q_s^{(\rm s)}$ the size-independent Binder ratio at the special transition point. We find $J_s^*=1.6222(10)$, where the number in parenthesis indicates the statistical uncertainty. This value agrees well with the estimates obtained by crossings. Furthermore, we obtain $y_{t1}^{(\rm s)}=0.61(2)$ and $Q_s^{(\rm s)}=0.843(2)$, both of which are in good agreement with the value 0.608 (4) and 0.840(1) of the 3D $XY$ model \cite{PhysRevE.72.016128}. 

\begin{figure}[!h]
\includegraphics[width=0.49 \columnwidth]{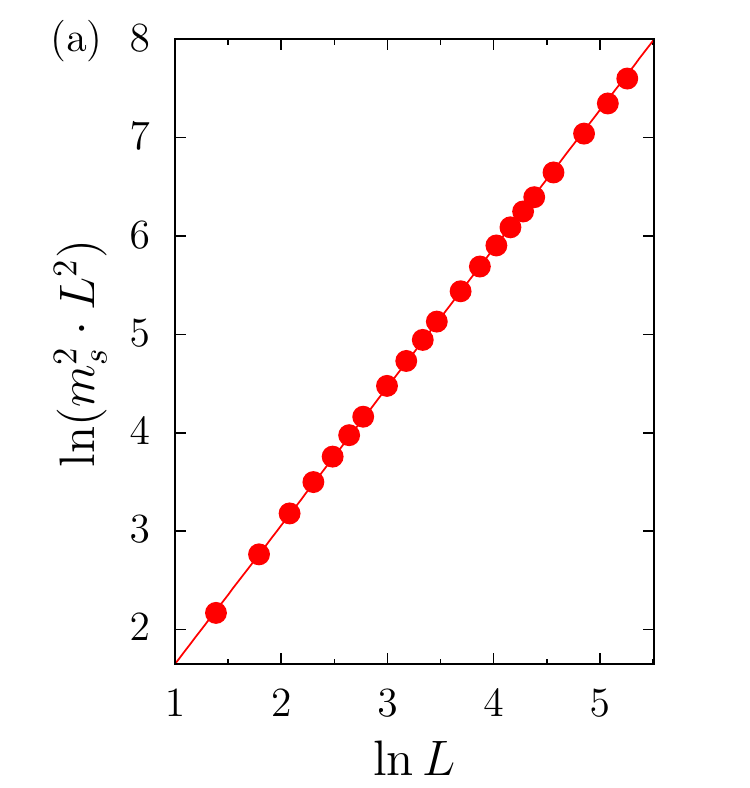}
\hspace{-0.45cm}
\includegraphics[width=0.49 \columnwidth]{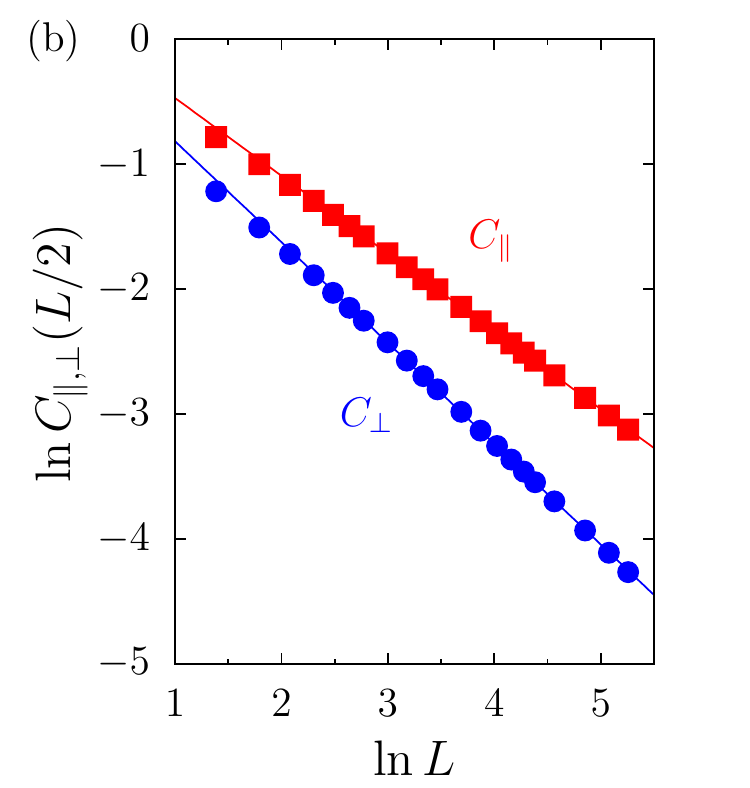}
	\caption{(a), $m_s^2L^2$ and (b), correlation $C_{\parallel, \perp}$ at the special transition point $J_s^*$. The largest size is $L=192$.
\label{special}}
\end{figure}

Exactly at $J^*_s=1.622$,  the surface criticality should be a special transition separating two surface transitions.  We expect
	\begin{equation}
		m_{s}^2 L^2 = a L^{2 y^{(\rm s)}_{h1}-2},
		\label{ms_spec}
	\end{equation}
in which $y_{h1}^{(\rm s)}$ is the scaling dimension of $h_1$ \cite{PhysRevB.9.2194, PhysRevE.72.016128}. This is evident from the simulation results shown in Fig. \ref{special}(a). The estimated scaling dimension is $y_{h1}^{(\rm s)} = 1.688(1)$, which is close to the scaling dimension $y_{h1}^{(\rm s)}=1.675(1)$ of the 3D $XY$  model \cite{PhysRevE.72.016128}. 

We have also calculated $C_\parallel(L/2)$ and $C_\perp(L/2)$ at $J^*_s=1.622$. The results are plotted in Fig. \ref{special}(b).
The finite-size behaviors of $C_\parallel(L/2)$ and $C_\perp(L/2)$  are
expected as the following forms
\begin{equation}
	C_\parallel(L/2) = a L^{-(1+\eta^{(\rm s)}_\parallel)},
	\label{cspec1}
\end{equation}
and
\begin{equation}
	C_\perp(L/2) = a L^{-(1+\eta^{(\rm s)}_\perp)},
	\label{cspec2}
\end{equation}
Fitting these scaling forms to our data,
we obtain $\eta^{(\rm s)}_\parallel=-0.372(3)$ and $\eta_\perp^{(\rm s)}=-0.184(4)$. 
For the reader's sake, we list these exponents in Tab.
\ref{tab_cri}.
We see that the exponents found for the special transition obey the relations in Eqs. (\ref{sc1}) and (\ref{sc2}) roughly.

\begin{table}[!h]
    \caption{Surface critical exponents at special transition. The exponents of the 3D $XY$ model \cite{PhysRevE.72.016128} are also listed for comparison.}
    \begin{tabular}{cccccc}
        \hline
        \hline
        Type      &   $J_s^* $  &    $y^{(\rm s)}_{t1}$   &     $y^{(\rm s)}_{h1}$     & $\eta^{(\rm s)}_\parallel$ & $\eta^{(\rm s)}_{\perp}$   \\
        \hline
        % 3D clock  &    1.622(1) &    0.55(1)          &      1.692(1)          & -0.377(2)              &  -0.158(1)             \\
        3D clock  &    1.622(1) &    0.61(2)          &      1.688(1)          & -0.372(3)              &  -0.184(4)             \\
        3D $XY$     &             &    0.608(4)         &       1.675(1)         &                        &                        \\
        \hline
        \hline
    \end{tabular}
    \label{tab_cri}
\end{table}

%%%%%%%%%%%%%%%%%%%%%%%%%%%%%%%%%%%%%%%%%%%%%%%%%%%%%%%%%%%%%%%%%%%%%%%%%%%%%%%%%%%%%%%%%%%%%%%%%%%%%%%%%%%%%%%%%%%%%%%%%

\subsection{Extraordinary-log transition}
\label{subsec:extra-log}
\begin{figure}[h]
\includegraphics[width=0.95 \columnwidth]{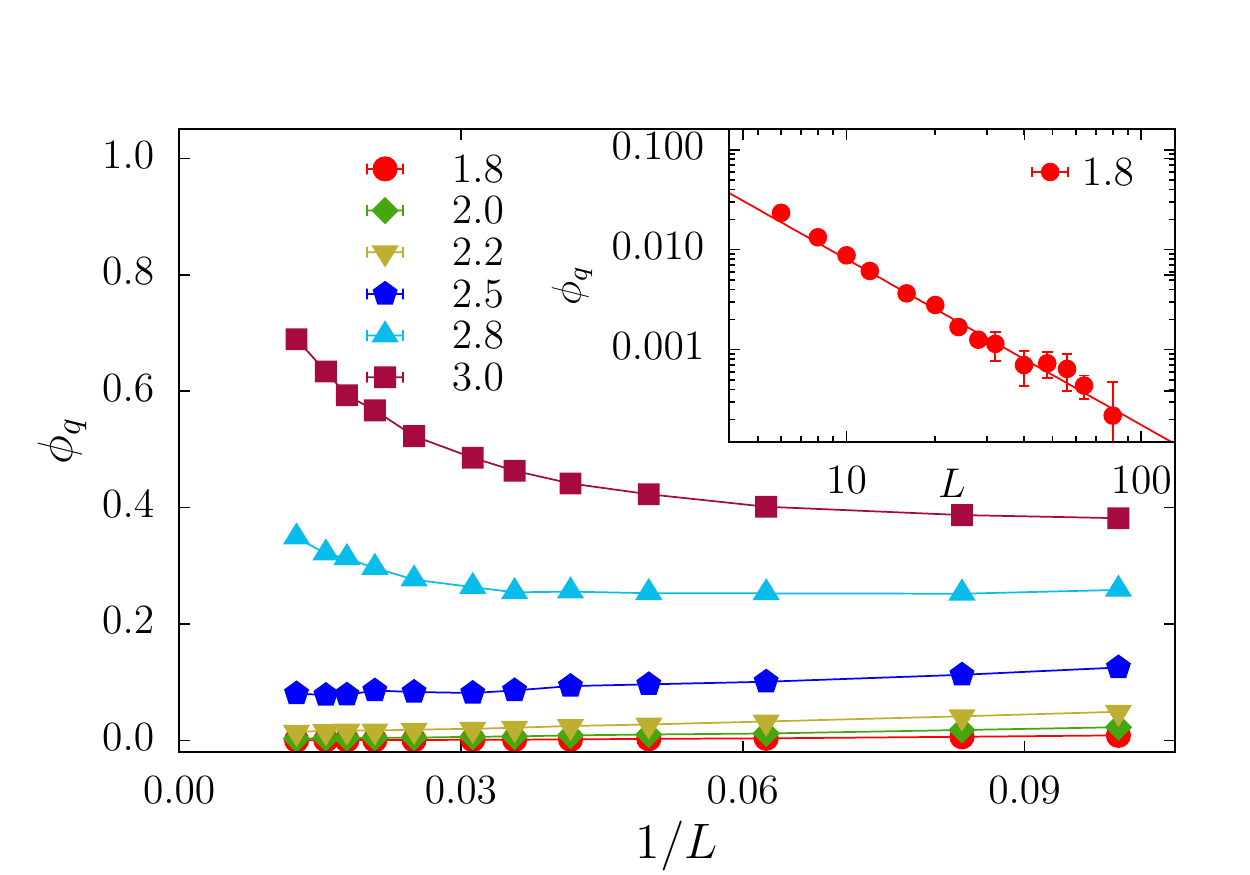}
\caption{$\phi_q$ at different $J_s$. When $J_{s}$ is small but still larger than $J_{s}^{*}$, $\phi_{q}$ flows to $0$ as the system size grows to infinity corresponding to the emergent O(2) symmetry.} 
\label{Fig:phiq}
\end{figure}

%Besides small $J_{s}$ and large $J_{s}$ limits, the central challenge is whether there is an intermediate phase between ordinary and extraordinary phases. 
We now  explore the symmetry of the surface phase when $J_{s} > J_{s}^{*}$. For a given spin configuration, we compute 
\begin{equation}
    M_{x}=\sum_{i \in {\rm surface}} \cos(\theta_i),
\end{equation} and 
\be
    M_{y}=\sum_{i \in {\rm surface}}\sin(\theta_{i}).
\ee 
With $M=(M_{x}^{2}+M_{y}^{2})^{1/2}$ and $\Theta = \arccos(M_{x}/M)$, 
we define an angular order parameter \cite{ShaoPRL} 
\begin{eqnarray}
\phi_q = \langle \cos(q\Theta)\rangle,
\end{eqnarray}
which becomes non-zero in response to the $Z_{q}$ symmetry. 
We have done simulations to calculate $\phi_q$ for several $J_s>J_s^*$. The results are shown in Fig. \ref{Fig:phiq}.

As discussed above, when $J_{s}$ is large enough, the relatively weak interaction between surface and bulk decouples the surface from the bulk, and the surface can be effectively described by the 2D $q$-state clock model. 
At the bulk critical temperature $T_{c}$, the surface enters the long-range ordered phase with $Z_{q}$ discrete symmetry spontaneous breaking. All the spins align to the same direction with $\theta = \frac{n2\pi}{q} (n=1,...,q)$. $\phi_q$ will flow to 1 with the system size increasing; when we decrease $J_{s}$ that is still larger than $J_{s}^{*}$, the flow has a totally different behavior as shown in Fig. \ref{Fig:phiq}. 

We find $\phi_q$ flows to zero  for $J_s$ up to $J_s \approx 2.3$. In $L \rightarrow \infty$ limit, $\phi_q \rightarrow 0$ indicates an emergent O(2) symmetry.
The decay of $\phi_q$ is analyzed according to following scaling form \cite{ShaoPRL}
\begin{equation}
    \phi_q \propto L^{y_q},
    \label{phiq}
\end{equation}
where $y_q$ is the scaling dimension of the $Z_q$ field. When $y_q<0$, the
field renormalized to zero at thermal dynamic limit \cite{ShaoPRL}.
% In $L \rightarrow \infty$ limit, $\phi_q \rightarrow 0$ which indicates an emergent $U(1)$ symmetry, i.e., there may be an intermediate extraordinary-log phase between disordered ordinary phase and long-range ordered extraordinary phase. 
Fitting according to Eq. (\ref{phiq}), we find $y_q$ for several values of $J_s$, as listed in Tab. \ref{eta-spin-correlation}. For $J_s=2.2$, the power $y_q$ is very small. To obtain a reasonable good fit of Eq. (\ref{phiq}), we have to use data of system size larger than $L=40$. 
We also notice that $y_q$ changes with $J_s$. This indicates the O(2) symmetry will finally reduce to $Z_q$ for large enough $J_s$.  

Recent theoretical research \cite{metlitski2020boundary} has predicted 
that there is an extraordinary-log transition at the bulk critical 
point of the $O(N)$ model when $N<N_{c}$, where $N_{c}$ is the critical dimension. This prediction has been verified numerically in a classical 3D $\phi^4$ model, which is an O(3) model \cite{PhysRevLett.126.135701}, and 3D $XY$ model \cite{PhysRevLett.127.120603}.
The emergent O(2) symmetry found in the surface of current model suggests that there may be an intermediate extraordinary-log phase characterized by the order parameter correlation decaying as a power of log$(r)$. We will show convincing numerical results below which indicate the existence of such an extraordinary-log phase.

In the extraordinary-log universality  class, the surface spin-spin correlation $C_{\parallel}(L/2)$ behaves as:
\be
C_{\parallel}(L/2) \sim [\ln(L/L_{0})]^{-\eta},
\label{clogL}
\ee
where $L_{0}$ is a non-universal constant. 

\begin{figure}[!h]
\includegraphics[width=0.49 \columnwidth]{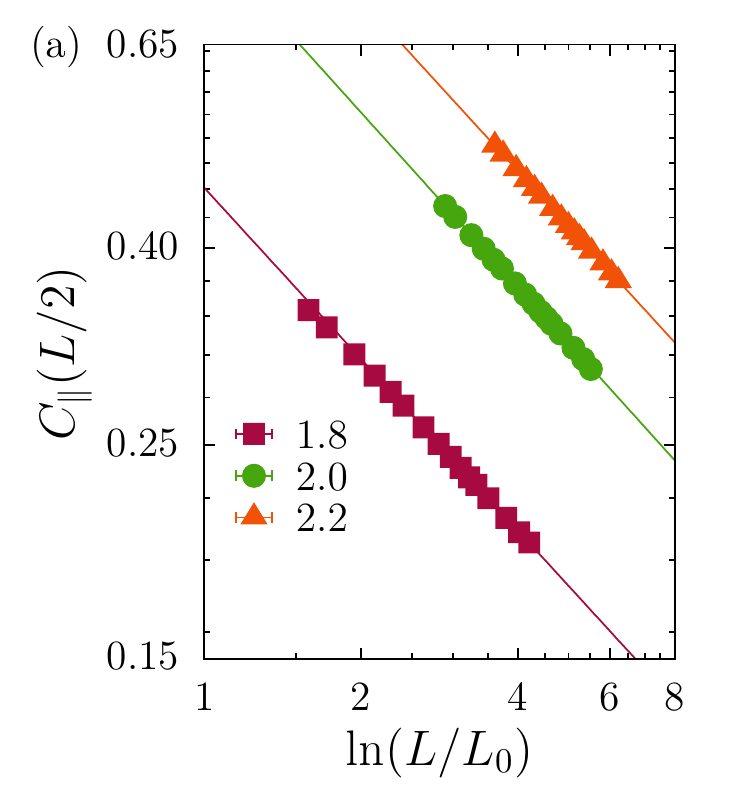}
\hspace{-0.45cm}
\includegraphics[width=0.49 \columnwidth]{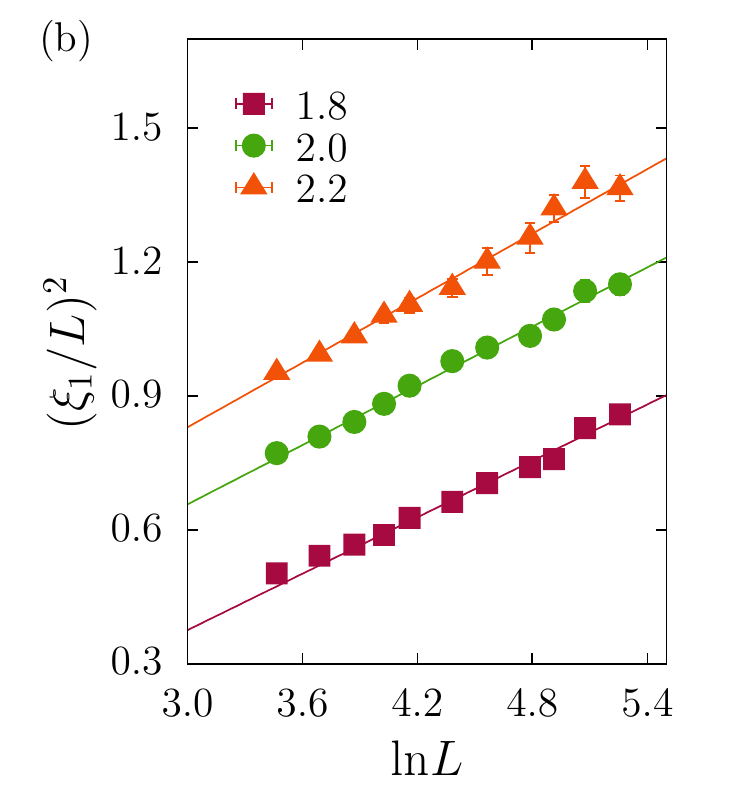}
	\caption{(a), $C_{\parallel}(L/2)$ and (b), $(\xi_1/L)^2$ for the extraordinary-log transition at $J_s=1.8, 2.0, 2.2$. The largest size is $L=192.$
\label{extra_log}}
\end{figure}

We have simulated systems at several $J_s$ up to size $L=192$. 
The results of $C_{\parallel}(L/2)$ for $J_s=1.8, 2.0$ and $2.2$  are shown in Fig. \ref{extra_log}(a).
Fitting according to Eq. (\ref{clogL}), we find statistical sound estimation of $\eta$ for $J_{s}=1.8, 2.0$ and $2.2$, as listed in 
Tab. \ref{eta-spin-correlation}. 
More details of the fits are listed in Tab.\ref{Cparallel_log}.
The critical exponent $\eta$ %estimated from $C_{\parallel}(L/2)$ 
fits well with the value 0.59 found in the 3D $XY$ model \cite{PhysRevLett.127.120603}. We have also tried to fit these data of $C_{\parallel}(L/2)$ in the conventional power law form. 
However, the exponents obtained keep changing as we discard small system sizes (See more details in the Appendix Tab. \ref{Cparallel_powerlaw}).

For large $J_s$, e.g., $J_s=2.8$, we see fits according to Eq. (\ref{clogL}) with different
minimal size $L_{\rm min}$ lead to drifting exponents, meanwhile, fits according to the conventional power law becomes stable, manifesting the expected extraordinary behavior. See Tab. \ref{Cparallel_log} and \ref{Cparallel_powerlaw}. This suggests a second special transition
has occurred.

\begin{table}[h]
\caption{The exponents $y_q$, $\eta$, $\alpha$  estimated for various surface coupling $J_s$ in the extraordinary-log phase.}
%\note{fill in $y_q$ here: $y_q$ changes with $J_s$}}
\begin{tabular}{cccc}
\hline
\hline
$J_s$    &  1.8       & 2.0              & 2.2          \\
\hline
$y_q$    &  -1.7(1)  & -1.0(1)         & -0.27(10)     \\
$\eta$   &  0.59(1)   & 0.60(3)          & 0.59(3)      \\   
$\alpha$ &  0.26(2)   & 0.24(4)          & 0.30(3)      \\      
\hline
\hline
\end{tabular}
\label{eta-spin-correlation}
\end{table}

In addition, in the extraordinary-log phase, the scaled surface second-moment correlation length $\xi_{1}/L$ 
%in a recent work on the SCBs of an improved model in the classical 3D O(3) universality class
%where the extraordinary-log phase is verified, 
is further proposed to scale as \cite{PhysRevLett.126.135701}:
\be
(\xi_{1}/L)^{2} \sim \frac{\alpha}{(N-1)} \ln(L/L^{\prime}) ,
\label{xi1log}
\ee
where $L^{\prime}$ is an unknown number. 
$\xi_{1}$ is  defined as:
\be
\xi_{1} = \frac{1}{2 \sin(\pi/L)} \sqrt{\frac{S(0)}{S(2\pi/L)}-1},
\ee
where the surface spin structure factor $S(k)$ is the Fourier transform of the spin-spin correlation function. 
We have also calculated $\xi_1$ for system size up to $L=192$ for several values of $J_s$. 
The results of $(\xi_1/L)^{2}$ as function of system size $L$ 
are shown in 
Fig. \ref{extra_log}(b).
Our results fits well with the scaling form Eq. (\ref{xi1log}). The best estimates of $\alpha$ for various surface coupling $J_{s}$ are listed 
in Tab. \ref{eta-spin-correlation}.  More details of the fits are shown in Tab. \ref{Rs_log} in the Appendix. The values of $\alpha$ estimated numerically are close to the value 0.27(2) found in  \cite{PhysRevLett.127.120603}. 

According to \cite{metlitski2020boundary}, the exponent $\eta$ which characterizes the finite size scaling of the spin-spin correlation is related to the RG parameter $\alpha$ in the form reads
\be
\eta = \frac{N-1}{2\pi \alpha},
\label{eta-alpha}
\ee
with $N=2$ in our model. Our numerical results agree well with this prediction.

Basing on these results, we conclude the existence of an extraordinary-log
phase in between of the ordinary transition and the extraordinary transition. The boundary (the first special point) between the ordinary phase and the extraordinary-log phase has been determined by using the surface magnetic
Binder ratio. The boundary between the extraordinary-log phase and the
extraordinary phase should be the point (the second special point)  where the $Z_q$ anisotropy becomes
relevant. At present work, we have not determined this point accurately.
Although the accurate second special point has not been determined, the 
above convincing numerical results provide strong evidence for the emergence of the extraordinary-log universality  class and intermediate extraordinary-log phase.

%%%%%%%%%%%%%%%%%%%%%%%%%%%%%%%%%%%%%%%%%%%%%%%%%%%%%%%%%%%%%%%%%%%%%%%%%%%%%%%%%%%%%%%%%%%%%%%%%%%%%%%%%%%%%%%%%%%%%%%%%

\section{Conclusions and discussions}
\label{conclusions}
In conclusion, we have studied the surface critical behaviors of the 3D 6-state clock model on the simple cubic lattice. The model has emergent O(2) symmetry at its bulk critical point, therefore, the transition is in the 3D O(2) universality class.
We obtained the schematic phase diagram of the model. We found that, when $J_{s} \approx J$, the model undergoes an ordinary transition
at the bulk critical point, which is in the SCB class of the 3D O(2) model, whereas, at $J_{s} \gg J$, the surface is described by the 2D $q$-state clock model and enters the $Z_{q}$ symmetry broken phase at bulk critical point. As a result, the surface undergoes an extraordinary transition with
$Z_q$ order presenting. This is different from the 3D O(2) model.  
We have also determined the two `melting' temperatures $T_{1}$ and $T_{2}$
corresponding to the two BKT transitions of the 2D clock model as functions of 
$J_{s} \gg J$, supporting the SCBs at these $J_s$ are of 
the extraordinary type. 

We found a special transition at $J_{s}^{*}=1.622(1)$ where the surface
leaving the ordinary region at bulk critical point. This is also the point the
higher BKT transition of the 2D clock model merges with the bulk $T_c$. We showed 
that this special point is in the special SCB of the 3D O(2) university class. 

We further showed that the emerged O(2) symmetry keeps to $J_s^\dagger>J_s>J_s^*$  by studying the scaling behavior the angular order parameter $\phi_q$ as the system size. We have found $J_s^\dagger >2.2$, but have not determined 
the exact value of it. As a 
result of this emergent O(2) symmetry in the surface, we 
proved numerically that
there is an intermediate extraordinary-log phase between the ordinary phase and the extraordinary phase. We found that the phase is controlled by the extraordinary-log fixed point of the 3D O(2) model
%and belongs to the logarithmic surface universality class
\cite{metlitski2020boundary} by showing the critical exponent $\eta$ 
and the RG parameter $\alpha$ determined in the 3D 6-state clock 
model agree well with the values of the 3D $XY$ model \cite{PhysRevLett.127.120603}.

%%%%%%%%%%%%%%%%%%%%%%%%%%%%%%%%%%%%%%%%%%%%%%%%%%%%%%%%%%%%%%%%%%%%%%%%%%%%%%%%%%%%%%%%%%%%%%%%%%%%%%%%%%%%%%%%%%%%%%%%%

\begin{acknowledgments}

W.G. were supported by the National Natural Science Foundation of China under Grant No.~12175015 and No.~11734002.
The authors acknowledge support extended by the Super Computing Center of Beijing Normal University.
\end{acknowledgments}

%%%%%%%%%%%%%%%%%%%%%%%%%%%%%%%%%%%%%%%%%%%%%%%%%%%%%%%%%%%%%%%%%%%%%%%%%%%%%%%%%%%%%%%%%%%%%%%%%%%%%%%%%%%%%%%%%%%%%%%%%

% \section{Appendix}

%%%%%%%%%%%%%%%%%%%%%%%%%%%%%%%%%%%%%%%%%%%%%%%%%%%%%%%%%%%%%%%%%%%%%%%%%%%%%%%%%%%%%%%%%%%%%%%%%%%%%%%%%%%%%%%%%%%%%%%%%

\bibliography{ref}

\begin{widetext}
\appendix
	%\section{Appendix\textbf{}}
	\section{Fitting details}

In this appendix, we show the fitting details of finite-size scaling analyses of the extraordinary-log and the extraordinary transitions. 

In the extraordinary-log phase, the spin-spin correlation $C_{\parallel}(L/2)$ behaves as:
\be
C_{\parallel}(L/2) = A [\ln(L/L_{0})]^{-\eta},
\label{eq_log}
\ee
and $\xi_{1}/L$ scales as:
\be
(\xi_{1}/L)^{2} =\alpha \ln L+B+\frac{C}{L} .
\label{eq_log_xi}
\ee

In the conventional extraordinary phase, the spin-spin correlation $C_{\parallel}(L/2)$ behaves as:
\be
	C_{\parallel}(L/2)=AL^{-(1+\eta_\parallel^{(\rm e)})}+C.
\ee

When $J_s$ is beyond the special point, where the emergent O(2) symmetry keeps, 
the decay of the spin-spin correlation $C_{\parallel}(L/2)$ is analyzed according to a power of 
	log($L$) 
as Eq. (\ref{eq_log}). The fitting results are shown in Tab. \ref{Cparallel_log}. The critical exponent $\eta$ is around $0.59$ when $1.622<J_s<2.4$, which agrees well with the previous results in the 3D $XY$ model \cite{PhysRevLett.127.120603} and indicates the existence of the extraordinary-log phase. 
When $J_{s} > 2.4$, the extracted values of $\eta$ deviate from 0.59.

Tab. \ref{Cparallel_powerlaw} shows the power law fitting results of the spin-spin correlation $C_{\parallel}(L/2)$. The exponents obtained keep changing as we discard small system sizes in the region of $J_s$ a bit larger than the special transition point, and become more stable for large $J_s$.

The scaling form of $\xi_1/L$ is shown in Eq. (\ref{eq_log_xi}) and the critical exponents $\alpha$ are listed in Tab. \ref{Rs_log}. The values of $\alpha$ estimated numerically are close to the value 0.27(2) found in 3D $XY$ model \cite{PhysRevLett.127.120603}.
%%%%%%%%%%%%%%%%%%%%%%%%%%%%%%%%%%%%%%%%%%%%%%%%%%%%%%%%%%%%%%%%%%%%%%%%%%%%%%%%%%%%%%%%%%%%%%%%%%%%%%%%%%%%%%%%%%%%%%%%%
% Table generated by Excel2LaTeX from sheet 'Sheet2'
\begin{table}[htbp]
  \centering
  \caption{Fits of $C_{\parallel}(L/2)=A[\ln(L/L_0)]^{-\eta}$}
    \begin{tabular}{cccccc}
    \toprule
    $J_s$    & $L_{\text{min}}$  & $\chi^2$/DOF & $A$     & $L_0$     & $\eta$ \\
    \hline
    1.8   & 8     & 58.875 & 0.75(4) & 1.02(10) & 0.81(3) \\
          & 16    & 6.022 & 0.54(1) & 1.9(1) & 0.67(1) \\
          & 32    & 1.212 & 0.46(1) & 2.9(2) & 0.59(1) \\
          & 48    & 1.498 & 0.45(2) & 3.1(4) & 0.58(3) \\
          & 64    & 2.035 & 0.46(7) & 2.9(12) & 0.59(7) \\
          & 72    & 2.711 & 0.5(1) & 2.8(19) & 0.59(11) \\
    2.0   & 8     & 3.960 & 0.89(1) & 0.65(3) & 0.627(7) \\
          & 16    & 1.819 & 0.82(2) & 0.81(6) & 0.591(12) \\
          & 32    & 1.721 & 0.84(7) & 0.8(2) & 0.60(3) \\
          & 48    & 1.778 & 0.9(2) & 0.6(4) & 0.63(8) \\
          & 64    & 2.700 & 0.8(4) & 0.8(10) & 0.60(17) \\
          & 72    & 1.638 & 0.8(4) & 0.8(13) & 0.60(20) \\
    2.2   & 8     & 1.760 & 1.13(2) & 0.34(1) & 0.605(6) \\
          & 16    & 1.208 & 1.07(2) & 0.40(3) & 0.584(9) \\
          & 32    & 1.376 & 1.09(8) & 0.38(9) & 0.59(3) \\
          & 48    & 1.535 & 1.1(2) & 0.3(2) & 0.61(7) \\
          & 64    & 2.189 & 1.1(4) & 0.4(5) & 0.59(14) \\
    2.4   & 8     & 1.631 & 1.33(2) & 0.200(8) & 0.596(5) \\
          & 16    & 1.607 & 1.33(4) & 0.20(2) & 0.59(1) \\
          & 32    & 1.171 & 1.24(7) & 0.26(5) & 0.57(2) \\
          & 48    & 1.458 & 1.2(2) & 0.3(1) & 0.56(5) \\
          & 64    & 1.305 & 1.0(2) & 0.6(5) & 0.50(8) \\
          & 72    & 1.790 & 1.1(5) & 0.5(8) & 0.52(16) \\
    2.5   & 8     & 1.123 & 1.34(2) & 0.188(7) & 0.568(5) \\
          & 16    & 0.665 & 1.37(3) & 0.17(1) & 0.576(9) \\
          & 32    & 0.788 & 1.3(1) & 0.19(5) & 0.57(3) \\
          & 48    & 0.791 & 1.3(2) & 0.2(1) & 0.57(6) \\
          & 64    & 0.502 & 1.4(4) & 0.2(2) & 0.59(9) \\
          & 72    & 0.477 & 2.1(14) & 0.03(8) & 0.7(2) \\
          & 80    & 0.336 & 1.3(6) & 0.2(4) & 0.55(16) \\
    2.6   & 8     & 1.794 & 1.31(2) & 0.195(9) & 0.530(5) \\
          & 16    & 1.970 & 1.26(4) & 0.23(3) & 0.52(1) \\
          & 32    & 1.023 & 1.04(5) & 0.49(9) & 0.44(2) \\
          & 48    & 0.727 & 0.93(6) & 0.8(2) & 0.40(3) \\
          & 64    & 1.032 & 0.9(1) & 1.0(6) & 0.39(5) \\
          & 72    & 0.770 & 1.1(3) & 0.3(4) & 0.47(10) \\
          & 80    & 0.909 & 1.4(8) & 0.1(3) & 0.54(19) \\
    2.8   & 8     & 43.955 & 0.99(3) & 0.51(7) & 0.357(13) \\
          & 16    & 20.068 & 0.85(3) & 1.1(2) & 0.292(14) \\
          & 32    & 4.217 & 0.71(1) & 3.4(5) & 0.209(10) \\
          & 48    & 2.916 & 0.66(2) & 5.7(11) & 0.175(14) \\
          & 64    & 1.766 & 0.62(2) & 9.9(22) & 0.141(14) \\
          & 72    & 0.732 & 0.59(1) & 15.1(25) & 0.116(10) \\
    \hline
    \end{tabular}%
  \label{Cparallel_log}%
\end{table}%

\begin{table}[htbp]
  \centering
  \caption{Fits of $C_{\parallel}(L/2)=AL^{-\eta}+C$}
    \begin{tabular}{cccccc}
    \toprule
    $J_s$    & $L_{\text{min}}$  & $\chi^2$/DOF & $A$     & $\eta$     & $C$ \\
    \hline
    1.8   & 8     & 9.957 & 0.826(5) & 0.557(4) & 0.1547(9) \\
          & 16    & 6.834 & 0.80(1) & 0.538(7) & 0.152(1) \\
          & 32    & 2.519 & 0.73(2) & 0.504(9) & 0.146(2) \\
          & 48    & 1.509 & 0.69(2) & 0.48(1) & 0.142(2) \\
          & 64    & 1.636 & 0.64(5) & 0.44(3) & 0.137(5) \\
          & 72    & 2.120 & 0.62(7) & 0.44(4) & 0.135(8) \\
          & 80    & 0.543 & 0.73(6) & 0.50(3) & 0.144(4) \\
          & 96    & 0.803 & 0.8(2) & 0.53(7) & 0.148(7) \\
    2.0   & 8     & 17.703 & 0.658(5) & 0.450(7) & 0.242(2) \\
          & 16    & 7.834 & 0.620(8) & 0.41(1) & 0.229(4) \\
          & 32    & 2.474 & 0.58(1) & 0.36(1) & 0.213(5) \\
          & 48    & 1.995 & 0.56(2) & 0.33(3) & 0.20(1) \\
          & 64    & 1.809 & 0.51(2) & 0.26(5) & 0.17(3) \\
          & 72    & 1.782 & 0.55(6) & 0.32(7) & 0.20(3) \\
          & 80    & 0.785 & 0.50(1) & 0.20(7) & 0.13(7) \\
    2.2   & 8     & 11.808 & 0.595(2) & 0.368(6) & 0.286(3) \\
          & 16    & 4.294 & 0.581(3) & 0.340(6) & 0.274(3) \\
          & 32    & 1.831 & 0.559(6) & 0.31(1) & 0.258(5) \\
          & 48    & 1.627 & 0.55(1) & 0.29(2) & 0.25(1) \\
          & 64    & 2.404 & 0.55(3) & 0.29(5) & 0.25(2) \\
          & 72    & 2.289 & 0.52(2) & 0.24(6) & 0.22(5) \\
          & 80    & 2.404 & 0.52(4) & 0.18(9) & 0.2(1) \\
    2.6   & 8     & 6.542 & 0.5496(8) & 0.311(4) & 0.366(2) \\
          & 16    & 0.938 & 0.5452(7) & 0.291(3) & 0.354(2) \\
          & 32    & 0.683 & 0.548(3) & 0.295(7) & 0.356(4) \\
          & 48    & 0.800 & 0.552(9) & 0.30(1) & 0.360(7) \\
          & 64    & 1.190 & 0.55(2) & 0.30(3) & 0.36(1) \\
          & 72    & 0.801 & 0.52(1) & 0.25(4) & 0.33(2) \\
          & 80    & 0.905 & 0.513(8) & 0.22(5) & 0.31(4) \\
          & 96    & 1.715 & 0.51(1) & 0.2(1) & 0.3(1) \\
    2.8   & 8     & 7.053 & 0.505(1) & 0.352(4) & 0.450(2) \\
          & 16    & 6.344 & 0.512(4) & 0.366(8) & 0.455(3) \\
          & 32    & 2.098 & 0.56(1) & 0.42(1) & 0.470(3) \\
          & 48    & 1.741 & 0.59(3) & 0.45(2) & 0.475(4) \\
          & 64    & 1.298 & 0.69(7) & 0.51(4) & 0.484(5) \\
          & 72    & 0.658 & 0.9(1) & 0.59(4) & 0.492(4) \\
          & 80    & 0.978 & 0.8(2) & 0.58(8) & 0.492(7) \\
          & 96    & 1.640 & 1.1(9) & 0.7(2) & 0.50(2) \\
    \hline
    \end{tabular}%
  \label{Cparallel_powerlaw}%
\end{table}%

%%%%%%%%%%%%%%%%%%%%%%%%%%%%%%%%%%%%%%%%%%%%%%%%%%%%%%%%%%%%%%%%%%%%%%%%%%%%%%%%%%%%%%%%%%%%%%%%%%%%%%%%%%%%%%%%%%%%%%%%%

%%%%%%%%%%%%%%%%%%%%%%%%%%%%%%%%%%%%%%%%%%%%%%%%%%%%%%%%%%%%%%%%%%%%%%%%%%%%%%%%%%%%%%%%%%%%%%%%%%%%%%%%%%%%%%%%%%%%%%%%%

% Table generated by Excel2LaTeX from sheet 'Sheet3'
\begin{table}[htbp]
  \centering
  \caption{Fits of $(\xi_1/L)^2=\alpha\ln L+B+C/L$}
    \begin{tabular}{cccccc}
    \toprule
    $J_s$    & $L_{\text{min}}$  & $\chi^2$/DOF & $\alpha$     & $B$     & $C$ \\
    1.8   & 16    & 0.952 & 0.251(7) & -0.48(3) & 4.1(2) \\
          & 20    & 0.829 & 0.260(8) & -0.53(4) & 4.5(3) \\
          & 24    & 0.873 & 0.264(10) & -0.55(5) & 4.7(5) \\
          & 28    & 0.884 & 0.257(12) & -0.52(6) & 4.3(7) \\
          & 32    & 0.776 & 0.244(15) & -0.44(8) & 3.3(9) \\
          & 40    & 0.760 & 0.26(2) & -0.5(1) & 4.3(13) \\
          & 48    & 0.839 & 0.25(3) & -0.5(2) & 3.4(24) \\
          & 56    & 0.977 & 0.25(4) & -0.5(2) & 3.7(42) \\
          & 64    & 0.937 & 0.28(5) & -0.7(3) & 7.4(53) \\
    2.0   & 16    & 1.235 & 0.284(12) & -0.33(6) & 3.6(4) \\
          & 20    & 0.993 & 0.26(2) & -0.19(8) & 2.3(8) \\
          & 24    & 1.002 & 0.25(2) & -0.14(10) & 1.7(10) \\
          & 28    & 0.663 & 0.21(2) & 0.03(11) & -0.4(11) \\
          & 32    & 0.547 & 0.23(2) & -0.07(11) & 1.1(13) \\
          & 40    & 0.537 & 0.21(3) & 0.05(16) & -0.9(23) \\
          & 48    & 0.603 & 0.20(4) & 0.10(22) & -1.7(34) \\
          & 56    & 0.635 & 0.22(5) & -0.02(27) & 0.6(46) \\
          & 64    & 0.589 & 0.26(6) & -0.26(33) & 5.6(60) \\
    2.2   & 16    & 0.506 & 0.295(12) & -0.18(6) & 3.5(4) \\
          & 20    & 0.432 & 0.312(15) & -0.26(7) & 4.4(6) \\
          & 24    & 0.456 & 0.32(2) & -0.30(9) & 4.8(9) \\
          & 28    & 0.506 & 0.32(2) & -0.29(13) & 4.7(14) \\
          & 32    & 0.540 & 0.31(3) & -0.2(2) & 3.9(19) \\
          & 40    & 0.607 & 0.29(5) & -0.2(3) & 2.9(35) \\
          & 48    & 0.708 & 0.30(7) & -0.2(4) & 3.1(59) \\
    \hline
    \end{tabular}%
  \label{Rs_log}%
\end{table}%

%%%%%%%%%%%%%%%%%%%%%%%%%%%%%%%%%%%%%%%%%%%%%%%%%%%%%%%%%%%%%%%%%%%%%%%%%%%%%%%%%%%%%%%%%%%%%%%%%%%%%%%%%%%%%%%%%%%%%%%%%
\end{widetext}
\end{CJK*}
\end{document}